%% file: root.tex
\documentclass[journal, 10 pt, twoside,web]{ieeecolor}
\usepackage{lcsys}
\def\BibTeX{{\rm B\kern-.05em{\sc i\kern-.025em b}\kern-.08em
    T\kern-.1667em\lower.7ex\hbox{E}\kern-.125emX}}
\usepackage{cite}
\usepackage{amsmath,amssymb,amsfonts}
\usepackage{algorithmic}
\usepackage{graphicx}
\usepackage{textcomp}
\usepackage{xcolor}
\usepackage{stmaryrd}
\usepackage{mathtools}
\usepackage{xcolor, soul}
\usepackage{nicematrix}
\usepackage{hyperref}

\newtheorem{theorem}{Theorem}
\newtheorem{corollary}{Corollary}

\newtheorem{lemma}{Lemma}
\newtheorem{remark}{Remark}
\newtheorem{proposition}{Proposition}
\newtheorem{defn}{Definition}

\newcommand{\Reals}{\mathbb{R}}

\newcommand{\id}{\textbf{I}}
\newcommand{\zeros}{\mathbf{0}}
\newcommand{\ones}{\mathbf{1}}
\newcommand{\norm}[1]{\left\lVert #1 \right\rVert}
\newcommand{\abs}[1]{\left\lvert #1 \right\rvert}

\newcommand{\oneToN}[1]{\{1,\dots,#1\}}
\newcommand{\closedInt}[2]{\llbracket #1,#2 \rrbracket}

\renewcommand{\st}{\text{ s.t. }}
\newcommand{\conv}[1]{\textbf{conv}\!\left(#1\right)}
\newcommand{\cone}[1]{\textbf{cone}\!\left(#1\right)}
\newcommand{\relaxation}[1]{\textbf{relax}\!\left(#1\right)}
\newcommand{\Binf}{\mathcal{B}_\infty}

\newcommand{\CZshort}[4]{\left\langle #1,#2,#3,#4 \right\rangle}
\newcommand{\HZshort}[6]{\left\langle #1,#2,#3,#4,#5,#6 \right\rangle}
\newcommand{\HZshortZO}[6]{\left\langle #1,#2,#3,#4,#5,#6 \right\rangle_{01}}

\newcommand{\J}{\mathcal{J}}
\newcommand{\N}{\mathcal{N}}
\newcommand{\X}{\mathcal{X}}

\newcommand{\Z}{\mathcal{Z}}
\newcommand{\U}{\mathcal{U}}
\renewcommand{\H}{\mathcal{H}}
\renewcommand{\L}{\mathcal{L}}

\newcommand{\revres}{}

\begin{document}

\title{Sharp Hybrid Zonotopes: Set Operations and the Reformulation-linearization Technique}
\author{Jonah J. Glunt, Joshua A. Robbins, \revres{Jacob A. Siefert}, Daniel Silvestre, and Herschel C. Pangborn
\thanks{Jonah J. Glunt, Joshua A. Robbins, \revres{Jacob A. Siefert,} and Herschel C. Pangborn are with The Pennsylvania State University, University Park, PA 16802 USA (e-mail: jglunt@psu.edu; jrobbins@psu.edu; \revres{jas7031@psu.edu;} hcpangborn@psu.edu).}
\thanks{Daniel Silvestre is with the School of Science and Technology, NOVA University of Lisbon, 1049-001 Lisbon, Portugal, also with COPELABS, Lusófona University, 1749-024 Lisbon, Portugal, and also with the Institute for Systems and Robotics, Instituto Superior Técnico, University of Lisbon, 1649-004 Lisbon, Portugal (e-mail: dsilvestre@isr.tecnico.ulisboa.pt).}
\thanks{This material is based upon work supported by the U.S. Department of Defense through the National Defense Science \& Engineering Graduate (NDSEG) Fellowship Program. The work from D. Silvestre was funded by FCT – Fundação para a Ciência e a Tecnologia through CTS - Centro de Tecnologia e Sistemas/UNINOVA/FCT/NOVA with reference CTS/00066.}
}

\maketitle
\thispagestyle{empty}

\begin{abstract}
\input{0_Abstract}
\end{abstract}


\input{1_Intro}
\input{2_Prelims}
\input{2_c_RLT}
\input{3_Sharp_Unions}
\input{4_RLT_with_HZ}
\input{5_Example}
\input{6_Conclusion}

\bibliographystyle{ieeetr}
\bibliography{bib}
\end{document}

%% file: 0_Abstract.tex
Mixed integer set representations, and specifically hybrid zonotopes, have enabled new techniques for reachability and verification of nonlinear and hybrid systems. Mixed-integer sets which have the property that their convex relaxation is equal to their convex hull are said to be \emph{sharp}. This property allows the convex hull to be computed with minimal overhead, and is known to be important for improving the convergence rates of mixed-integer optimization algorithms that rely on convex relaxations.
This paper examines methods for formulating sharp hybrid zonotopes and provides sharpness-preserving methods for performing several key set operations. The paper then shows how the \emph{reformulation-linearization technique} can be applied to create a sharp realization of a hybrid zonotope that is initially not sharp. A numerical example applies this technique to find the convex hull of a level set of a feedforward ReLU neural network.

%% file: 1_Intro.tex
\section{Introduction}

\IEEEPARstart{H}{ybrid} zonotopes have been applied for reachability analysis, verification, and planning in a variety of applications, including hybrid systems~\cite{SiefertHybSUS, Hadjiloizou2024}, model predictive control~\cite{Bird_HybZonoMPC}, neural networks~\cite{Zhang2023,Zhang2024,Ortiz2023_HZ_ReLU,Siefert_Reach_FuncDecomp}, and autonomous vehicles~\cite{Robbins2024_ACC_EnergyAware, Robbins2024_MotionPlanner_TightHZ}. However, the computational cost of solving mixed-integer problems can be untenable in some cases. Mitigating strategies include the use of convex relaxations, as in the context of spacecraft with thruster deadbands~\cite{accikmecse2011lossless}, or an iterative procedure combining convex relations and variable instantiation~\cite{deadbandThrusters}.

The \emph{convex relaxation} (or simply \emph{relaxation}, for brevity) of a mixed-integer set is the set obtained by replacing integrality constraints with their interval hulls.
A set is said to be sharp if its convex relaxation is its convex hull. 
To the best of the authors' knowledge, the first formal definition of a set representation being sharp comes from~\cite{Jeroslow1984_MIPR}, which based the idea on the \emph{linear relaxation optimal} mixed-integer programming models from~\cite{Meyer1981}. Ref.~\cite{Balas1985_DisjunctiveProgramming} is another fundamental work which examines the relaxations of disjunctive programs and can, for example, be used to construct sharp representations of the union of polyhedra. 

Formulating mixed-integer optimization programs whose constraints define sharp sets
has been shown to reduce the number of branch-and-bound iterations, as less conservative objective function bounds reduce the number of branches that need to be checked~\cite{Frangioni2006,Gunluk2008,Belotti2016, Marcucci2024_GCS,Wu2025}. 
Solution strategies for mixed-integer optimal and model predictive control have found reduced computational burden by utilizing formulations that exploit sharp relaxations~\cite{Axehill2010, Marcucci2024_GCS}. 
Specific mixed-integer problems that have benefited from tighter (if not sharp) relaxations include optimal planning and scheduling of a batch processing plant~\cite{ORCUN1999}, optimal control of pulse width modulation inputs~\cite{Schmitt2013}, gas pipeline network control~\cite{Wu2017}, verification of image classification networks~\cite{Anderson2020}, reachability of neural feedback systems~\cite{Zhang2023a}, robot manipulation~\cite{Graesdal2024}, and motion planning~\cite{Marcucci2024_Thesis,Robbins2024_MotionPlanner_TightHZ}. While this paper focuses on mixed-integer representations, the same ideas of sharpness hold for purely integer programs and sets~\cite{Fiorini2021}.

Techniques have been proposed for reformulating a given set representation into a sharp one without any specific knowledge of \revres{how the set was obtained}~\cite{Sherali1994,Lovasz1991,Lasserre2001}. This paper is concerned with the \emph{reformulation-linearization technique} (RLT) of~\cite{Sherali1994} and its application to hybrid zonotopes. All these methods experience worst-case exponential growth in the complexity of the sharp representations, although they do have approximate forms or order reduction methods to mitigate the growth of the set description. 

Alongside reformulation approaches, one can also utilize set constructions and operations that inherently induce or preserve sharpness. For mixed-integer sets \revres{(of which hybrid zonotopes are a subset)},~\cite{Jeroslow1988} examines the relaxations of various set operations, and demonstrates that Cartesian products and affine transformations preserve sharpness. Ref.~\cite{Jeroslow1988} also provides a formulation for the union of mixed-integer sets that preserves sharpness, \revres{which we not only specify to hybrid zonotopes, but also expand upon by providing geometric intuition for performing unions with hybrid zonotopes.}

\textit{Contributions:}
This paper studies the formulation of sharp hybrid zonotopes and preservation of sharpness through key hybrid zonotope set operations. Specifically, the paper contributes a novel identity for the union of hybrid zonotopes that preserves sharpness and proves that affine mappings and Cartesian products also preserve sharpness of hybrid zonotopes. Additionally, this is the first paper to apply the RLT to hybrid zonotopes, and we demonstrate its utility in producing sharp versions as well as convex hulls of hybrid zonotopes. 

%% file: 2_Prelims.tex
\section{Preliminaries}
\label{sec:2_Prelims}

\subsection{Notation}
Matrices are denoted by uppercase letters, e.g., $G\in\Reals^{n\times n_g}$, and sets by uppercase calligraphic letters, e.g., $\Z\subset\Reals^{n}$. 
Vectors and scalars are denoted by lowercase letters. 
The $i^{th}$ column of a matrix $G$ is denoted $G_i$.
Matrices of all $0$ and $1$ elements are denoted by $\zeros$ and $\ones$ (respectively) \revres{and $\id$ denotes the identity matrix}, with dimension noted in subscript when not clear from context.
The closed interval $\{x\in\Reals^n \mid a \leq x \leq b\}$ is notated $\closedInt{a}{b}$. 
The cardinality of a set $\J$ is denoted $\abs{\J}$. The convex hull of a set $\mathcal{X}$ is denoted $\conv{\mathcal{X}}$, \revres{and the conic hull by $\cone{\X}$}.
The $n$-dimensional unit hypercube is $\Binf^n=\{x\in\Reals^n\mid\norm{x}_\infty\leq 1\}$.

\subsection{Set Representations and Operations}
\begin{defn}\cite[Defn. 3]{scott2016constrained}
    The set $\mathcal{Z}_c \subset \Reals^n$ is a \emph{constrained zonotope} if there exist $G\in\Reals^{n\times n_g}$, $c\in\Reals^n$, $A\in\Reals^{n_c \times n_g}$, and $b\in\Reals^{n_c}$ such that
    \begin{equation}\label{eqn:CZ_defn}
        \mathcal{Z}_c = \left\{ G\xi + c \mid A\xi=b,\; \xi \in\Binf^{n_g} \right\}.
    \end{equation}%
    $\Z_c$ is denoted in shorthand by $\CZshort{G}{c}{A}{b}$, and its complexity is denoted by the tuple $(n_g, n_c)$. Constrained zonotopes represent convex polytopes. 
\end{defn}
\begin{defn}\cite[Defn. 3]{Bird_HybZono}
    The set $\mathcal{Z}_h\subset\Reals^n$ is a \emph{hybrid zonotope} if there exist $G_c\in\Reals^{n\times n_{g}}$, $G_b\in\Reals^{n\times n_{b}}$, $c\in\Reals^{n}$, $A_c\in\Reals^{n_{c}\times n_{g}}$, $A_b\in\Reals^{n_{c}\times n_{b}}$, and $b\in\Reals^{n_c}$ such that 
    \begin{align}
        \Z_h = \big\{ G_c \xi_c + G_b \xi_b + c \mid\ &A_c \xi_c + A_b \xi_b = b\;, \nonumber\\
        &\!\!\xi_c\in\Binf^{n_g},\ \xi_b\in\{-1,1\}^{n_b}\big\}.\label{eqn:HZ_defn}
    \end{align}
    $\Z_h$ is denoted in shorthand by $\HZshort{G_c}{G_b}{c}{A_c}{A_b}{b}$, and its complexity is denoted by the tuple $(n_g, n_b, n_c)$. A hybrid zonotope is the union of $2^{n_b}$ constrained zonotopes corresponding to the possible combinations of binary factors, $\xi_b$.
\end{defn}
\begin{defn}
    The \emph{factor space} of a hybrid zonotope is the set of all continuous and binary factors $(\xi_c,\xi_b)$ that satisfy the constraints in~\eqref{eqn:HZ_defn}.
\end{defn}
\begin{remark}
    Ref.~\cite[Prop. 1]{Robbins2024_MotionPlanner_TightHZ} proved that hybrid zonotopes can be equivalently defined with factors $\xi_c\in\closedInt{0}{1}^{n_g}$ and $\xi_b \in \{0,1\}^{n_b}$ which we refer to as a \emph{01-hybrid zonotope}. This form will be used occasionally throughout this paper, as it sometimes leads to simpler proofs. This will be clearly specified in the text, and denoted in shorthand by $\Z = \HZshortZO{G_c}{G_b}{c}{A_c}{A_b}{b}$. 
\end{remark}

Identities for common set operations applied to hybrid zonotopes (along with the corresponding complexity analysis) can be found in~\cite{Bird_HybZono, BIRDthesis_2022} \revres{and are assumed to be used except when otherwise stated.} Hybrid zonotopes are implemented in the MATLAB toolbox zonoLAB\footnote{Code for the operations and example presented in this paper is also in zonoLAB; see \url{https://github.com/ESCL-at-UTD/zonoLAB}.}~\cite{Koeln2024_zonoLAB} along with tools for analysis of various linear, nonlinear, and hybrid systems. 
\begin{defn}
    Given a hybrid zonotope $\Z = \HZshort{G_c}{G_b}{c}{A_c}{A_b}{b}$, the \emph{convex relaxation} (or simply \emph{relaxation}) $\relaxation{\Z}$ is the constrained zonotope $\CZshort{[G_c\ G_b]}{c}{[A_c \ A_b]}{b}$. It is always true that $\Z\subseteq\conv{\Z}\subseteq\relaxation{\Z}$.
\end{defn}
\begin{defn}
    A hybrid zonotope $\Z$ is said to be \emph{sharp} if it has the property that $\relaxation{\Z}=\conv{\Z}$.
\end{defn}

\begin{remark}
    While the convex hull operator acts on sets, the convex relaxation is an operator on a set representation. Different representations of the same set can have different convex relaxations.
\end{remark}

%% file: 2_c_RLT.tex
\subsection{The Reformulation-linearization Technique}
The purpose of the reformulation-linearization technique (RLT) is to reformulate the constraints of a mixed-integer set (without any special knowledge of its structure) to yield a sharp representation. The RLT was first presented in~\cite{Sherali1992} for optimization problems and then later modified in~\cite{Sherali1994} for sets. 

Given the set $\X\subseteq\Reals^{n+m}$ represented as
\begin{align}
    \label{eqn:Sherali_X_defn}
    \X = \left\{ (x,y)\!\mid\! A x + B y = \beta, x \in\! \{0,1\}^n , y \in \closedInt{0}{1}^m \right\},
\end{align}
where $\beta\in\Reals^r$, the RLT constructs a hierarchy of sets $\X_d$ for $d \in \{1,\dots,n\}=\N$, which we review in this subsection.

\revres{The auxiliary variables $w$ are indexed using sets $\J\subseteq\N$ (instead of usual whole numbers), and the variables $v$ are indexed with both a set $\J\subseteq\N$ and an integer $k\in\{1,\dots,m\}$. For a given $\J\subseteq\{1,\dots,n\}$ and $k\in\{1,\dots,m\}$, then $w_\J, v_{\J,k}\in\Reals$. Define $w_\emptyset=1$, $w_{\{i\}}=x_i$ for $i\in\{1,\dots,n\}$, and $v_{\emptyset,k}=y_k$ for $k\in\{1,\dots,m\}$.}

Then, define the functions
\begin{align}
    f_d(\J_1,\J_2) &= \sum_{\emptyset\subseteq \mathcal{I} \subseteq \J_2} w_{\J_1 \cup \mathcal{I}} (-1)^{\abs{\mathcal{I}}}\;,\\
    \label{eqn:Sherali_fdk_defn}
    f_d^k(\J_1,\J_2) &= \sum_{\emptyset \subseteq \mathcal{I} \subseteq \J_2} v_{\J_1 \cup \mathcal{I},k}(-1)^{\abs{\mathcal{I}}}\;.
\end{align}%
Given sets of indices $\J_1,\J_2\subseteq\N$, we say that $(\J_1,\J_2)$ is of order $d$ if
\begin{equation}
    \J_1 \cap \J_2 = \emptyset \text{ and } \abs{\J_1 \cup \J_2}=d\;.
\end{equation}

Selecting $d\in\N$, then $\X_d$ is defined as
\begin{align}
    \X_d = \Bigg\{ (x,y,v,w) \Bigg\vert \left(\sum_{j\in \J} A_j-\beta \right) w_\J  + \sum_{j \in \N\setminus \J} A_j w_{\J\cup \{j\}} \nonumber \\
    + \sum_{k=1}^m B_k v_{\J,k} =0\  \forall\ \J\subseteq \N \st \abs{\J}\leq d\;, \nonumber \\
    f_D(\J_1,\J_2)\geq0\ \forall (\J_1,\J_2) \text{ of order } D=\min\{d+1,n\}\;, \nonumber\\
    f_d(\J_1,\J_2)\geq f_d^k(\J_1,\J_2)\geq0\ \forall k\in\oneToN{m}  \nonumber\\
    \label{eqn:Sherali_Xd_Defn}
    \text{ and } \forall (\J_1,\J_2) \text{ of order } d \Bigg\}\;.
\end{align}%

The projection of $\X_d$ onto the $(x,y)$ space is 
\begin{equation}
    \label{eqn:Sherali_XPd_defn}
    \X_{P,d} = \{(x,y) \mid \exists (w,v) \st (x,y,w,v) \in \X_d \}\;.
\end{equation}

\revres{Theorem~\ref{thm:Sherali_hier_equiv} establishes a set containment hierarchy for the previously defined relaxations and, importantly, proves that each of the relaxations exactly represents the original set $\X$ if binary restrictions on variables $x$ are reintroduced.}
\begin{theorem}\cite[Thm. 3.2]{Sherali1994}
\label{thm:Sherali_hier_equiv}
    Letting $\X_{P,d}$ denote the projected relaxation of $\X$ as defined by~\eqref{eqn:Sherali_Xd_Defn} and~\eqref{eqn:Sherali_XPd_defn} for each $d=1,\dots,n$, then
    \begin{equation}
        \conv{\X} \subseteq \X_{P,n} \subseteq \X_{P,n-1} \subseteq\! \cdots\! \subseteq \X_{P,1}\subseteq \relaxation{\X},
    \end{equation}
    \begin{equation}
        \X_{P,d}\cap \{(x,y) \mid x \in \{0,1\}^n\} = \X \ \forall d\in\oneToN{n}\;.
    \end{equation}
\end{theorem}

\begin{theorem}\cite[Thm. 3.5]{Sherali1994}
\label{thm:Sherali_cvxHull}
    Letting $\X_{P,n}$ denote the projected relaxation of $\X$ as defined by~\eqref{eqn:Sherali_Xd_Defn} and~\eqref{eqn:Sherali_XPd_defn} with $d=n$, then $\X_{P,n} = \conv{\X}$.
\end{theorem}

Theorem~\ref{thm:Sherali_cvxHull} proves that the $n$th relaxation is equivalent to the convex hull of the set (the tightest possible convex relaxation). Defining $\mathcal{X}' = \revres{\X_{P,n}}\cap\{(x,y)\mid x\in\{0,1\}^n\}$, then it is true that $\X'=\X$ and $\relaxation{\X'} = \conv{\X}$, hence $\X'$ is a sharp representation of the original set $\X$.

%% file: 3_Sharp_Unions.tex
\section{Hybrid Zonotope Operations That Preserve Sharpness}
\label{sec:sharp_HZ_ops}
Here we provide methods for constructing sharp hybrid zonotopes. \revres{We review simple theoretical results for Minkowski sum, affine mapping, and Cartesian product, then use those results to present a novel identity for the sharpness-preserving union of hybrid zonotopes.} These operations can be used to manipulate hybrid zonotopes while preserving sharpness, or to construct sharp hybrid zonotopes via set operations on constrained zonotopes (treated as hybrid zonotopes with no binary generators). Previous work provided a sharp hybrid zonotope identity for the union of vertex-representation polytopes~\cite[Thm. 5]{Siefert_Reach_FuncDecomp},~\cite[Thm. 3.3]{Robbins2024_MotionPlanner_TightHZ}, which is another useful construction for sharp hybrid zonotopes. 

\subsection{Operations on Hybrid Zonotopes}

\begin{theorem}\cite[Lemma 2]{Robbins2024_MotionPlanner_TightHZ}\label{thm:mink-sum-is-sharp}
    Given sharp hybrid zonotopes $\mathcal{Z}_1$ and $\mathcal{Z}_2$, the Minkowski sum $\mathcal{Z}_1 \oplus \mathcal{Z}_2$ is sharp.
\end{theorem}

\begin{theorem} \label{thm:linmap-is-sharp}
    Given a sharp hybrid zonotope $\Z=\HZshort{G_c}{G_b}{c}{A_c}{A_b}{b}$ and affine mapping $T(z)=Rz+s$ for $z\in\Z$, then the image $T(\Z)=\HZshort{RG_c}{R G_b}{Rc+s}{A_c}{A_b}{b}$ is sharp.

    \begin{proof}
        Hybrid zonotopes are a subset of \emph{bounded-mixed-integer representable} sets as defined in \cite{Jeroslow1988}. For a bounded-mixed-integer representable set $\mathcal{S}$, the affine map $T(\mathcal{S})$ preserves sharpness~\cite[Lemma 3.4]{Jeroslow1988}.
    \end{proof}
\end{theorem}

\begin{theorem} \label{thm:cart-prod-is-sharp}
    Given sharp hybrid zonotopes $\mathcal{Z}_1=\left\langle G_{c,1},  G_{b,1}, c_1, A_{c,1}, A_{b,1}, b_1 \right\rangle$ and $\mathcal{Z}_2=\left\langle G_{c,2}, G_{b,2}, c_2, A_{c,2}, A_{b,2}, b_2 \right\rangle$, the Cartesian product $\mathcal{Z} = \mathcal{Z}_1 \times \mathcal{Z}_2$, defined as
    \begin{multline}
        \mathcal{Z} = \left\langle \begin{bmatrix} G_{c,1} & \zeros \\ \zeros & G_{c,2} \end{bmatrix}, \begin{bmatrix} G_{b,1} & \zeros \\ \zeros & G_{b,2} \end{bmatrix}, \begin{bmatrix} c_1 \\ c_2 \end{bmatrix}, \right. \\ \left. \begin{bmatrix} A_{c,1} & \zeros \\ \zeros & A_{c,2} \end{bmatrix}, \begin{bmatrix} A_{b,1} & \zeros \\ \zeros & A_{b,2} \end{bmatrix}, \begin{bmatrix} b_1 \\ b_2 `\end{bmatrix} \right\rangle\;,
    \end{multline} is sharp.

    \begin{proof}
        Taking the convex hull $\conv{\mathcal{Z}}$ and using the fact that the convex hull and Cartesian product commute yields $\conv{\mathcal{Z}} = \conv{\mathcal{Z}_1} \times \conv{\mathcal{Z}_2}$. Because $\mathcal{Z}_1$ and $\mathcal{Z}_2$ are sharp, $\conv{\mathcal{Z}} = \relaxation{\mathcal{Z}_1} \times \relaxation{\mathcal{Z}_2}$, and
        
        \begin{multline}
            \conv{\mathcal{Z}} = \left\langle \begin{bmatrix} 
        G_{c,1} & G_{b,1} & 0 & 0 \\
        0 & 0 & G_{c,2} & G_{b,2}
        \end{bmatrix}, \begin{bmatrix} c_1 \\ c_2 \end{bmatrix} \right. \\ \left.
        \begin{bmatrix} 
        A_{c,1} & A_{b,1} & 0 & 0 \\
        0 & 0 & A_{c,2} & A_{b,2}
        \end{bmatrix}, \begin{bmatrix} b_1 \\ b_2 \end{bmatrix} \right\rangle \;,
        \end{multline}
        which is equivalent to $\relaxation{\mathcal{Z}}$.
    \end{proof}
\end{theorem}

\subsection{Union of Hybrid Zonotopes}

\revres{We first review a result regarding the relaxation of intersections that will be used in later theorems. Proposition~\ref{prop:union_HZ_vec} provides an identity for the union of a hybrid zonotope with a single point, which is then shown to be sharp. Proposition~\ref{prop:union_of_many} uses that result to produce a general union of hybrid zonotopes, which is also shown to be sharp.}
\begin{lemma}\cite[Lemma 3.2 (unproved)]{Jeroslow1988}
    \label{thm:relax_of_intsct}
    \revres{
    Given hybrid zonotopes $\X,\Z\subseteq\Reals^n$, $\relaxation{\X\cap\Z} = \relaxation{\X} \cap \relaxation{\Z}$.}

    \begin{proof}
        \revres{Writing the matrices to represent the sets on both sides of the equality yields the same set representation.}
    \end{proof}
    
\end{lemma}

\begin{proposition}
    \label{prop:union_HZ_vec}
    \revres{
    Given $\Z = \HZshortZO{G_c}{G_b}{c}{A_c}{A_b}{b} \subset \Reals^n$ and $x\in\Reals^n$, then $\Z \cup \{x\}$ is given by $\Z'=\HZshortZO{G_c'}{G_b'}{c'}{A_c'}{A_b'}{b'}$, where
    \allowdisplaybreaks
    \begin{align}
        G_c' &= \begin{bmatrix} G_c & \zeros_{n\times(n_g+n_b)} \end{bmatrix}\;,\; G_b' = \begin{bmatrix} G_b & c-x \end{bmatrix}\;,\; c' = x\;,\nonumber\\
        A_c' &= \begin{bmatrix} A_c & \zeros_{n_c\times (n_g+n_b)} \\ \begin{bmatrix} \id_{n_g} \\ \zeros_{n_b \times n_g} \end{bmatrix} & \id_{n_g+n_b} \end{bmatrix}\;,\nonumber\\
        \label{eqn:union_HZ_vec}
        A_b' &= \begin{bmatrix} A_b & -b \\ \begin{bmatrix} \zeros_{n_g\times n_b} \\ \id_{n_b} \end{bmatrix} & -\ones_{(n_g+n_b)\times1} \end{bmatrix} ,\: b' = \zeros_{(n_c+1)\times1}\,,
    \end{align}
    and $(n_g, n_b, n_c)$ is the complexity of $\Z$. Further, if $\Z$ is sharp, then $\Z'$ is sharp.
    }

    \begin{proof}
        \revres{Let $\xi = (\xi_c, \xi_b)\in\Reals^{n_g+n_b}$ be the factors associated with $\Z$, and $s\in\closedInt{0}{1}^{n_g+n_b}$, $\sigma\in\{0,1\}$ the additional factors with $Z'$. Let $G = [G_c\ G_b]$ and $A = [A_c\ A_b]$. Then, expanding~\eqref{eqn:union_HZ_vec} yields that $z\in\Z'$ if and only if $z = G\xi + (c-x)\sigma + x$, $A\xi = \sigma b$, and $\xi_i + s_i = \sigma\ \forall i\in\{1,\dots,n_g+n_b\}$. Equivalently, one could define $\lambda_1 = \sigma$, $\lambda_2 = 1-\sigma$, and then $z\in \Z'$ if and only if $z = z_1+z_2$, $z_1 = G\xi + \lambda_1 c$, $A\xi = \lambda_1 b$, $z_2=\lambda_2 x$, $0\leq\xi\leq\ones\lambda_1$, $\lambda_1+\lambda_2=1$, which by~\cite[Thm. 3.1]{Jeroslow1988} defines a sharp representation of the union $\Z \cup \{x\}$.    }
    \end{proof}
\end{proposition}

\begin{corollary}
    \label{cor:complexity_union_HZ_vec}
    \revres{The complexity of the hybrid zonotope resulting from Proposition~\ref{prop:union_HZ_vec} is $(2n_g+n_b, n_b+1, n_g+n_b+n_c)$, where $(n_g,n_b,n_c)$ is the complexity of $\Z$. 
    }
\end{corollary}

\begin{proposition}
    \label{prop:union_of_many}
    \revres{Given the hybrid zonotopes $\Z_i\subset\Reals^n$ for $i\in\{1,\dots,N\}$ and letting $\U_i = (\Z_i \times \{1\}) \cup \{\zeros\}$, then
    \begin{align}
        \label{eqn:union_HZs_combo}
        \bigcup_{i=1}^N \Z_i &= \begin{bmatrix} \id_n & \zeros_{n\times1} \end{bmatrix} \left( \left( \bigoplus_{i=1}^N \mathcal{U}_i \right) \cap_{[\zeros_{1\times n} \; 1]} \{1\} \right)\;.
    \end{align}}

    \begin{proof}
        \revres{Expanding the definition of $\U_i$ yields
        \begin{equation}
            \mathcal{U}_i = \left\{ \begin{bmatrix} z_i \\ b_i \end{bmatrix} \middle| \begin{matrix}
                b_i \in \{0,1\} \;, \\
                z_i \in \begin{cases}
                    \Z_i\;, & \text{if } b_i = 1\;,\\
                    \{\zeros\}\;, & \text{else}\;.
                \end{cases}
            \end{matrix} \right\}\;.
        \end{equation}
        Thus, the right side of~\eqref{eqn:union_HZs_combo} gives
        \begin{equation}
            \left\{ \sum_{i=1}^N z_i \middle| \begin{matrix}
                b_i \in\{0,1\} \ \forall i\in\{1,\dots,N\}\;,\\
                \sum_{i=1}^N b_i = 1\;,\\
                z_i \in \begin{cases}
                    \Z_i\;, & \text{if } b_i = 1\;,\\
                    \{\zeros\}\;, & \text{else}\;.
                \end{cases}\ \forall i\in\{1,\dots,N\} 
            \end{matrix}\right\}\;.
        \end{equation}
        The constraint $\sum b_i = 1$ ensures that if $b_j=1$, then $b_i=0\ \forall i\neq j$. Thus, $\sum z_i = z_j$, and therefore a point $z$ is contained in the representation~\eqref{eqn:union_HZs_combo} of $\bigcup\Z_i$ if and only if $z = z_j \in \Z_j$ for some $j\in\{1,\dots,N\}$. }
    \end{proof}
\end{proposition}

\begin{figure}
    \centering
    \includegraphics[width=.75\linewidth, trim={.1in 0in .1in .44in},clip]{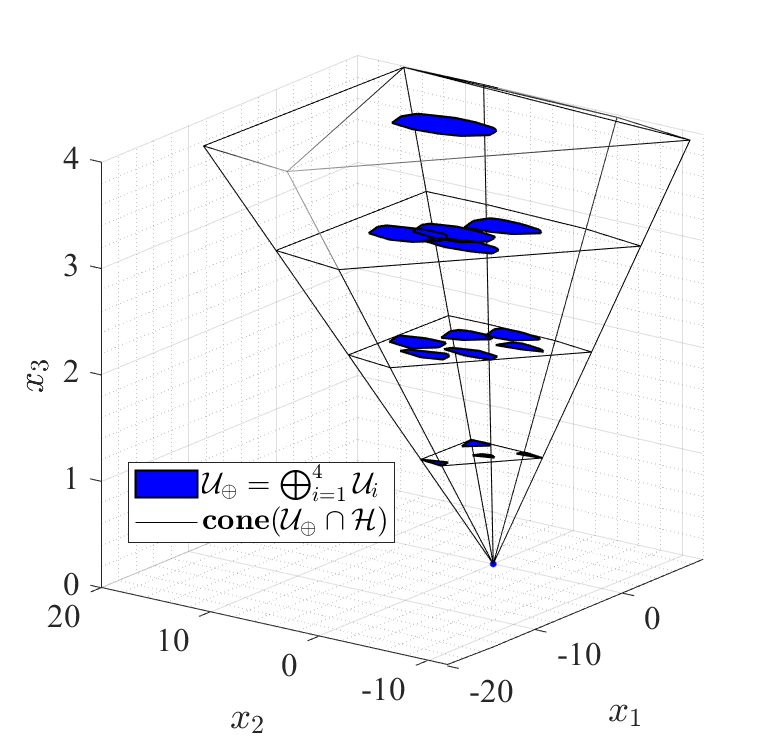}
    \caption{Applying Proposition~\ref{prop:union_of_many} to four polytopes yields the resulting set $\U_\oplus$ plotted in blue, which contains $\binom{4}{k}$ combinations at $x_3=k$ in the lifted space. Intersecting this set with the hyperplane $x_3=1$ yields the union of the four polytopes. As demonstrated in the proof of Theorem~\ref{thm:sharpness_union_of_many}, $\U_\oplus\subseteq\cone{\U_\oplus \cap \H}$, which is plotted in black wireframe.}
    \label{fig:U_oplus}
\end{figure}

\revres{Figure~\ref{fig:U_oplus} depicts the intermediate set $\U_\oplus = \bigoplus_{i=1}^4 \U_i$ of Proposition~\ref{prop:union_of_many} applied to four polytopes $\Z_i$. This set represents the $\binom{4}{0},\ \binom{4}{1}, \dots, \binom{4}{4}$ combinations for the Minkowski sum of $\Z_i$, and intersecting it with the hyperplane $x_3=1$ yields points which are in exactly one of the sets $\Z_i$.}

\begin{theorem}
    \label{thm:sharpness_union_of_many}
    
    \revres{When the union $(\Z_i\times\{1\}) \cup \{0\}$ is performed via Proposition~\ref{prop:union_HZ_vec} (or other sharpness-preserving identity for the union of a hybrid zonotope and a point), then Proposition~\ref{prop:union_of_many} preserves sharpness.}

    \begin{proof}
        \revres{Suppose that all $\Z_i$ for $i\in\{1,\dots,N\}$ are sharp, and by way of contradiction, suppose the set resulting from Proposition~\ref{prop:union_of_many} is not sharp. Let $\U_\oplus = \bigoplus_{i=1}^N \U_i$ and $\L=\U_\oplus\cap_{[\zeros\ 1]}\{1\} = \U_\oplus \cap \H$, where the hyperplane $\H=\Reals^n\times\{1\}$. Then, $\U_\oplus \subseteq \bigoplus_{i=1}^N \left( \bigcup_{k=1}^N \U_k \right)\subseteq \bigoplus_{i=1}^N \cone{ \bigcup_{k=1}^N \U_k } = \cone{\bigcup_{k=1}^N \U_k}$. Since $\bigcup_{k=1}^N \U_k = \left( \bigcup_{k=1}^N \Z_k \times\{1\} \right) \cup \{\zeros\}$ and $\cone{\bigcup_{k=1}^N \U_k}$ already includes the origin, then $\cone{\bigcup_{k=1}^N \U_k} = \cone{\bigcup_{k=1}^N \Z_k \times\{1\}}$. Because $\bigcup_{i=1}^N \Z_i = [\id\ \zeros] \L$ by~\eqref{eqn:union_HZs_combo}, and $[\zeros\ 1] \L = \{1\}$, then $\L = \bigcup_{i=1}^N \Z_i \times \{1\}$. Therefore, $\U_\oplus \subseteq \cone{\L}$.}
        
        \revres{$\U_\oplus$ is sharp, as all operations in its construction have sharpness-preserving identities. Since it is assumed $\L$ is not sharp, there exists $\ell\in\relaxation{\L}$ such that $\ell\notin\conv{\L}=\conv{\L}\cap\H$. By Lemma~\ref{thm:relax_of_intsct}, $\ell\in\relaxation{\U_\oplus}\cap\H = \conv{\U_\oplus}\cap\H = \cone{\L} \cap\H$. Because $\L=\L\cap\H$, then $\cone{\L}\cap\H = \conv{\L}$, hence $\ell\in\conv{\L}$.}
    \end{proof}
\end{theorem}

\begin{corollary}
    \label{cor:complexity_union_of_many}
    \revres{The union identity in Proposition~\ref{prop:union_of_many} (using Proposition~\ref{prop:union_HZ_vec} for union with a vector) generates hybrid zonotopes with complexity
    \allowdisplaybreaks
    \begin{align}
    n_g &= \sum_{i=1}^N \left(2n_{g,i}+n_{b,i}\right)\;,\; n_b = N + \sum_{i=1}^N n_{b,i}\;,\nonumber\\
    n_c &= 1 + \sum_{i=1}^N \left( n_{g,i}+n_{b,i}+n_{c,i} \right)\;,
    \end{align}
    where $(n_{g,i},n_{b,i},n_{c,i})$ is the complexity of $\Z_i$.}

    \begin{proof}
        \revres{From Corollary~\ref{cor:complexity_union_HZ_vec}, the complexity of each $\mathcal{U}_i$ is $(2n_{g,i}+n_{b,i}, n_{b,i}+1, n_{g,i}+n_{b,i}+n_{c,i})$. Upon taking the Minkowski sum in~\eqref{eqn:union_HZs_combo}, those complexities are summed, and the generalized intersection adds a single constraint.}
    \end{proof}
\end{corollary}

%% file: 4_RLT_with_HZ.tex
\section{Application of the RLT to Hybrid Zonotopes}
\label{sec:Application_to_HZs}

\revres{The methods presented in Section~\ref{sec:sharp_HZ_ops} can be used to construct sharp hybrid zonotopes by applying sharpness-preserving set operations to zonotopes, constrained zonotopes, or other sharp hybrid zonotopes. However when this is not possible, it is beneficial to have a method to produce a sharp representation of any hybrid zonotope. To that end, this section examines the application of the RLT to hybrid zonotopes.}

Comparing the definition of a hybrid zonotope~\eqref{eqn:HZ_defn} to the set definition~\eqref{eqn:Sherali_X_defn}, the right hand sides are equivalent after mapping the hybrid zonotope to a 01-hybrid zonotope. Then, the RLT can be carried out to write all of the constraints for the respective $\X_d$ in~\eqref{eqn:Sherali_Xd_Defn}. 

\begin{theorem}
    Applying the RLT to a 01-hybrid zonotope results in a set $\X_{P,d}$ that can also be represented by a 01-hybrid zonotope. 

\begin{proof}
If the RLT can be applied to make a sharp representation of the factor space of a hybrid zonotope, then by Theorem~\ref{thm:linmap-is-sharp}, the hybrid zonotope itself will be sharp. 
The affine equality constraints can be written in a hybrid zonotope format, and we now discuss how to add the appropriate slack variables for the inequalities. 

The constraints involving $f_d(\J_1,\J_2)$ and $f_d^k(\J_1,\J_2)$ in~\eqref{eqn:Sherali_Xd_Defn} arise in the RLT by substituting 
\begin{align}
    w_\J = \prod_{j\in\J} x_j\;,\; 
    v_{\J,k} = y_k \prod_{j\in\J} x_j\;,
\end{align}
into the product
\begin{align}
    \label{eqn:Sherali_Fd_defn}
    F_d(\J_1, \J_2) &= \left( \prod_{j\in \J_1} x_j \right) \left( \prod_{j\in \revres{\J_2}} (1-x_j) \right)\;.
\end{align}
Because $0\leq x_j \leq 1$, then $F_d(\J_1,\J_2)\geq0$ by definition~\eqref{eqn:Sherali_Fd_defn}. However, the linearization expression $f_d(\J_1,\J_2)$ is affine in $w$ with coefficients from $\{-1,1\}$, which is not inherently positive, hence the inclusion of this inequality in~\eqref{eqn:Sherali_Xd_Defn}. Similarly, $F_d(\J_1,\J_2)\leq 1$, so while the constraint $f_d(\J_1,\J_2)\leq1$ is not necessary, it is always valid. Therefore the inequality constraint $f_d(\J_1,\J_2)\geq0$ can be replaced with the equality constraint $f_d(\J_1,\J_2)=s_i$, where $s_i\in\closedInt{0}{1}$. The same process can be carried out for the double inequalities of the form $f_d(\J_1,\J_2)\geq f_d^k(\J_1,\J_2) \geq 0$. Therefore, all of the constraints for $\X_d$ can be written as affine equalities with variables from either $\{0,1\}$ or $\closedInt{0}{1}$, so $\X_d$ is a hybrid zonotope. As projections are affine mappings, and affine mappings preserve sharpness of hybrid zonotopes (Thm.~\ref{thm:linmap-is-sharp}), then $\X_{P,d}$ can also be written as a hybrid zonotope.
\end{proof}
\end{theorem}

Thus we have shown that the RLT can be applied to the factor space of a hybrid zonotope, and after performing the affine mapping of the factor space (where all new factors added during the RLT have generator vectors $\zeros_{n\times 1}$), we obtain a sharp representation of the original hybrid zonotope. At the expense of adding exponentially many new constraints and variables, the RLT can be used with hybrid zonotopes to 1) construct a sharp hybrid zonotope equivalent to a given hybrid zonotope, 2) improve the tightness of the relaxation of a given hybrid zonotope, and 3) construct the convex hull of a hybrid zonotope as a constrained zonotope. 

The parameter $d\in\{1,2,\dots,n_b\}$ can be used to tune the tradeoff between having a tight relaxation and set representation complexity. When $d$ is small, the relaxation is a larger set with fewer constraints and variables. When $d=n_b$, the relaxation is the convex hull and requires the most complexity.

\begin{corollary}
\label{cor:RLT_HZ_complexity}
Given a hybrid zonotope with complexity $(n_g, n_b, n_c)$, then the hybrid zonotope output of the RLT with $d\in\{1,\dots,n_b\}$ has complexity $(n_g', n_b', n_c')$, where
\label{eqn:complexity_of_RLT_HZs}
\begin{align}
    n_g' &= 2^{n_b}(n_g+1)+2^{d+1} \binom{n_b}{d} n_g - n_b - 1\;,\; n_b' = n_b\;,\nonumber\\
    n_c' &= n_c \sum_{i = 0}^d \binom{n_b}{i} + 2^{d+1} \binom{n_b}{d} n_g\;.
\end{align}

\begin{proof}
There are $2^{n_b}-n_g-1$ continuous variables $w_\J$ where $\emptyset\neq\J\subseteq N$ (the $n_g$ subtracted is for the original $x$ variables), $2^{n_b}n_g$ continuous variables $v_{\J,k}$ (this is including the original variables $y$), and $2^{d+1}\binom{n_b}{d} n_g$ continuous slack variables arising from the inequality constraints. There are $n_c\binom{n_b}{i}$ affine equality constraints for sets $\J$ of order $i$, and we must consider all orders from $\{0,\dots,d\}$. Further, there are the same $2^{d+1} \binom{n_b}{d} n_g$ inequalities that were counted for the slack variables. The number of binary variables is unchanged.
\end{proof}
\end{corollary}

When $d=n_b$, the RLT yields a sharp hybrid zonotope with
\begin{align}
\label{eqn:complexity_of_RLT_dn}
    n_g' &= 2^{n_b} (3n_g+1)-n_b-1 \;, n_b' = n_b \;,\nonumber\\
    n_c' &= 2^{n_b}(2n_g+n_c)\;.
\end{align}

%% file: 5_Example.tex
\section{Numerical Example}
\label{sec:example}

Consider a neural network trained to learn some cost metric over the state space; for example, this could be the probability of a location on a map being occupied by an obstacle or pedestrian. One approach for making occupancy decisions is to take level sets of the neural network function, thereby identifying all locations with a cost higher than some threshold to be flagged as potential ``obstacles." As the neural network output function is likely nonconvex, one may be interested in representing the convex hull of the obstacles to have a single polytope that needs to be avoided (instead of an implicitly defined mixed-integer set). 

\begin{figure}
    \centering
    \includegraphics[width=.9\linewidth]{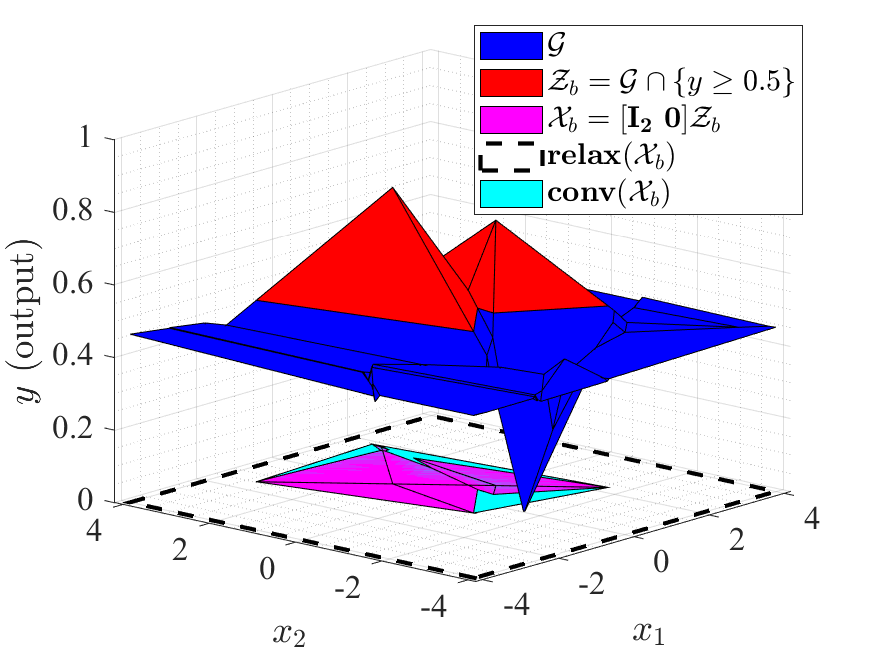}
    \caption{The set $\X_b$ (plotted in magenta) is a projection of a halfspace intersection of $\mathcal{G}$. $\X_b$ is not sharp, as its relaxation (plotted with the dashed line) can visually be seen to be much larger than $\conv{\X_b}$. Applying the RLT, we obtain a sharp representation, and plot its convex hull in cyan.}
    \label{fig:Ex_NN_badrelax}
\end{figure}

For this example, let us consider a feedforward ReLU neural network trained to learn the MATLAB function \texttt{peaks} over the domain $\closedInt{-4}{4}^2$, where the function has been normalized to the output domain $\closedInt{0}{1}$. The network has two hidden layers with ten nodes each and was trained for 1000 epochs on 10,000 uniformly distributed points. The graph of the neural network output function $\mathcal{G}=\{(x,N(x))\mid x\in\closedInt{-4}{4}^2\}$, where $N(x)$ is the neural network evaluated at input $x$, can be represented exactly as a hybrid zonotope~\cite{Ortiz2023_HZ_ReLU}, and is plotted in blue in~Figure~\ref{fig:Ex_NN_badrelax}. 

To find the unsafe states, we intersect the graph of the neural network with the halfspace $\{y\geq0.5\}$ to obtain the set $\Z_b = \{(x,N(x)) \mid x\in\closedInt{-4}{4}^2,\ N(x)\geq0.5\}$, plotted in red in Figure~\ref{fig:Ex_NN_badrelax}. $\Z_b$ can be projected down onto the state space to obtain the set $\X_b = [\id_2\ \zeros]\Z_b$, which represents all of the states which are unsafe. $\X_b$ is plotted in Figure~\ref{fig:Ex_NN_badrelax} in magenta. The set $\X_b$ has complexity $(83, 20, 61)$, which is reduced to $(21, 5, 14)$ by the redundancy removal technique in~\cite{BIRDthesis_2022}.

As is often the case after performing intersection operations, $\X_b$ is not sharp. This can be seen in Figure~\ref{fig:Ex_NN_badrelax} by the fact that $\relaxation{\X_b}$ (dashed line) is the entire domain. A sharp version of $\X_b$ with complexity $(2042, 5, 1792)$ is obtained via the RLT. Because $\X_b$ is sharp, its convex relaxation, a constrained zonotope with complexity $(2047, 1792)$, will be exactly its convex hull. \revres{Obtaining the exact convex hull of a hybrid zonotope was a previously unsolved problem without the RLT.} 

\begin{figure}
    \centering
    \includegraphics[width=.81\linewidth,trim={0 .45in 0 .5in},clip]{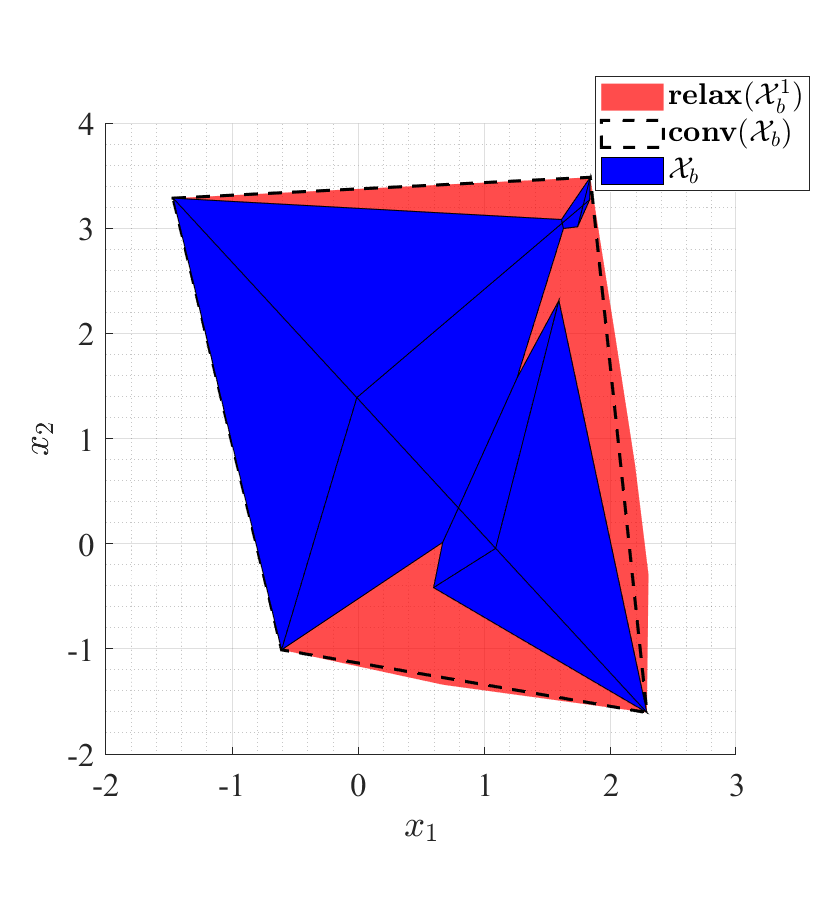}
    \caption{Performing the RLT with $d=1$ instead of $d=5$, we obtain the representation $\X_b^1$, whose relaxation is only slightly larger than the convex hull (but with a fraction of the complexity).}
    \label{fig:Ex_NN_relax_comp}
\end{figure}

If instead of the exact convex hull, we wanted an outer approximation (but with less complexity than running the full RLT), we could also perform the RLT with $d=1,2,3,4$ to obtain sets whose relaxations are somewhere between the whole state space and the convex hull. Computing the result for $d=1$ and calling it $\X_b^1$, we find the complexity of $\X_b^1$ is $(1118,5,504)$, and it is plotted in red in Figure~\ref{fig:Ex_NN_relax_comp}. We see that for less complexity, $\relaxation{\X_b^1}$ is a rather tight over-approximation of $\conv{\X_b}$; specifically, $\relaxation{\X_b^1}$ yields only a 3.2\% larger area than $\conv{\X_b}$. 

%% file: 6_Conclusion.tex
\section{Conclusion}
This paper presents a novel sharpness-preserving identity for the union of hybrid zonotopes, and proves that existing identities for affine mapping and Cartesian product preserve sharpness. Additionally, this paper applies the RLT to hybrid zonotopes, which has a parameter to tune the tradeoff between complexity and tightness of the relaxation. 
Future work will develop identities preserving sharpness for other set operations with hybrid zonotopes, as well as explore the use of sharp hybrid zonotopes within systems and controls optimization problems, such as model predictive control. 

%% file: root.bbl
\begin{thebibliography}{10}

\bibitem{SiefertHybSUS}
J.~A. Siefert, T.~J. Bird, J.~P. Koeln, N.~Jain, and H.~C. Pangborn, ``Robust successor and precursor sets of hybrid systems using hybrid zonotopes,'' {\em IEEE Control Systems Letters}, vol.~7, pp.~355--360, 2023.

\bibitem{Hadjiloizou2024}
L.~Hadjiloizou, F.~J. Jiang, A.~Alanwar, and K.~H. Johansson, ``Formal verification of linear temporal logic specifications using hybrid zonotope-based reachability analysis.'' European Control Conference, 2024.

\bibitem{Bird_HybZonoMPC}
T.~Bird, N.~Jain, H.~Pangborn, and J.~Koeln, ``Set-based reachability and the explicit solution of linear {MPC} using hybrid zonotopes.'' American Control Conference, 2022.

\bibitem{Zhang2023}
Y.~Zhang, H.~Zhang, and X.~Xu, ``Backward reachability analysis of neural feedback systems using hybrid zonotopes,'' {\em IEEE Control Systems Letters}, vol.~7, pp.~2779--2784, 2023.

\bibitem{Zhang2024}
Y.~Zhang, H.~Zhang, and X.~Xu, ``Reachability analysis of neural network control systems with tunable accuracy and efficiency,'' {\em IEEE Control Systems Letters}, vol.~8, pp.~1697--1702, 2024.

\bibitem{Ortiz2023_HZ_ReLU}
J.~Ortiz, A.~Vellucci, J.~Koeln, and J.~Ruths, ``Hybrid zonotopes exactly represent {ReLU} neural networks.'' IEEE Conference on Decision and Control, 2023.

\bibitem{Siefert_Reach_FuncDecomp}
J.~A. Siefert, T.~J. Bird, A.~F. Thompson, J.~J. Glunt, J.~P. Koeln, N.~Jain, and H.~C. Pangborn, ``Reachability analysis using hybrid zonotopes and functional decomposition,'' {\em Transactions on Automatic Control}, 2025.

\bibitem{Robbins2024_ACC_EnergyAware}
J.~A. Robbins, A.~F. Thompson, S.~Brennan, and H.~C. Pangborn, ``Energy-aware predictive motion planning for autonomous vehicles using a hybrid zonotope constraint representation.'' American Control Conference, 2025.

\bibitem{Robbins2024_MotionPlanner_TightHZ}
J.~A. Robbins, J.~A. Siefert, S.~Brennan, and H.~C. Pangborn, ``Mixed-integer {MPC}-based motion planning using hybrid zonotopes with tight relaxations.'' arXiv 2411.01286, 2024.

\bibitem{accikmecse2011lossless}
B.~A{\c{c}}{\i}kme{\c{s}}e and L.~Blackmore, ``Lossless convexification of a class of optimal control problems with non-convex control constraints,'' {\em Automatica}, vol.~47, no.~2, pp.~341--347, 2011.

\bibitem{deadbandThrusters}
P.~Taborda, H.~Matias, D.~Silvestre, and P.~Lourenço, ``Convex {MPC} and thrust allocation with deadband for spacecraft rendezvous,'' {\em IEEE Control Systems Letters}, vol.~8, pp.~1132--1137, 2024.

\bibitem{Jeroslow1984_MIPR}
R.~Jeroslow and J.~Lowe, ``Modeling with integer variables,'' {\em Mathematical Programming at Oberwolfach II}, vol.~22, pp.~167--184, 1984.

\bibitem{Meyer1981}
R.~R. Meyer, ``A theoretical and computational comparison of “equivalent” mixed-integer formulations,'' {\em Naval Research Logistics Quarterly}, vol.~28, pp.~115--131, 1981.

\bibitem{Balas1985_DisjunctiveProgramming}
E.~Balas, ``Disjunctive programming and a hierarchy of relaxations for discrete optimization problems,'' {\em SIAM Journal on Algebraic Discrete Methods}, vol.~6, 1985.

\bibitem{Frangioni2006}
A.~Frangioni and C.~Gentile, ``Perspective cuts for a class of convex 0-1 mixed integer programs,'' {\em Mathematical Programming}, vol.~106, pp.~225--236, 2006.

\bibitem{Gunluk2008}
O.~Günlük and J.~Linderoth, ``Perspective relaxation of mixed integer nonlinear programs with indicator variables.'' Integer Programming and Combinatorial Optimization, 2008.

\bibitem{Belotti2016}
P.~Belotti, P.~Bonami, M.~Fischetti, A.~Lodi, M.~Monaci, A.~Nogales-Gómez, and D.~Salvagnin, ``On handling indicator constraints in mixed integer programming,'' {\em Computational Optimization and Applications}, vol.~65, pp.~545--566, 2016.

\bibitem{Marcucci2024_GCS}
T.~Marcucci, J.~Umenberger, P.~Parrilo, and R.~Tedrake, ``Shortest paths in graphs of convex sets,'' {\em SIAM Journal on Optimization}, vol.~34, pp.~507--532, 2024.

\bibitem{Wu2025}
O.~Wu, P.~Muts, I.~Nowak, and E.~M.~T. Hendrix, ``On the use of overlapping convex hull relaxations to solve nonconvex {MINLPs},'' {\em Journal of Global Optimization}, vol.~91, no.~2, pp.~415--436, 2025.

\bibitem{Axehill2010}
D.~Axehill, L.~Vandenberghe, and A.~Hansson, ``Convex relaxations for mixed integer predictive control,'' {\em Automatica}, vol.~46, no.~9, pp.~1540--1545, 2010.

\bibitem{ORCUN1999}
S.~Or\c{c}un, {\.I}.~Altinel, and {\"O}.~Horta\c{c}su, ``Reducing the integrality gap with a modified re-formulation linearization approach,'' {\em Computers and Chemical Engineering}, vol.~23, pp.~539--542, 1999.

\bibitem{Schmitt2013}
M.~Schmitt, R.~Vujanic, J.~Warrington, and M.~Morari, ``An approach for model predictive control of mixed integer-input linear systems based on convex relaxations.'' IEEE Conference on Decision and Control, 2013.

\bibitem{Wu2017}
F.~Wu, H.~Nagarajan, A.~Zlotnik, R.~Sioshansi, and A.~M. Rudkevich, ``Adaptive convex relaxations for gas pipeline network optimization.'' American Control Conference, 2017.

\bibitem{Anderson2020}
R.~Anderson, J.~Huchette, W.~Ma, C.~Tjandraatmadja, and J.~P. Vielma, ``Strong mixed-integer programming formulations for trained neural networks,'' {\em Mathematical Programming}, vol.~183, no.~1, pp.~3--39, 2020.

\bibitem{Zhang2023a}
Y.~Zhang and X.~Xu, ``Reachability analysis and safety verification of neural feedback systems via hybrid zonotopes.'' American Control Conference, 2023.

\bibitem{Graesdal2024}
B.~P. Graesdal, S.~Y.~C. Chia, T.~Marcucci, S.~Morozov, A.~Amice, P.~A. Parrilo, and R.~Tedrake, ``Towards tight convex relaxations for contact-rich manipulation.'' Robotics: Science and Systems, 2024.

\bibitem{Marcucci2024_Thesis}
T.~Marcucci, ``Graphs of convex sets with applications to optimal control and motion planning.'' Massachusetts Institute of Technology, 2024.

\bibitem{Fiorini2021}
S.~Fiorini, T.~Huynh, and S.~Weltge, ``Strengthening convex relaxations of 0/1-sets using {Boolean} formulas,'' {\em Mathematical Programming}, vol.~190, pp.~467--482, 2021.

\bibitem{Sherali1994}
H.~D. Sherali and W.~P. Adams, ``A hierarchy of relaxations and convex hull characterizations for mixed-integer zero-one programming problems,'' {\em Discrete Applied Mathematics}, vol.~52, pp.~83--106, 1994.

\bibitem{Lovasz1991}
L.~{Lovász} and A.~Schrijver, ``Cones of matrices and set-functions and 0–1 optimization,'' {\em SIAM Journal on Optimization}, vol.~1, pp.~166--190, 1991.

\bibitem{Lasserre2001}
J.~B. Lasserre, ``An explicit exact {SDP} relaxation for nonlinear 0-1 programs.'' Integer Programming and Combinatorial Optimization, 2001.

\bibitem{Jeroslow1988}
R.~G. Jeroslow, ``Alternative formulations of mixed integer programs,'' {\em Annals of Operations Research}, vol.~12, pp.~241--276, 1988.

\bibitem{scott2016constrained}
J.~K. Scott, D.~M. Raimondo, G.~R. Marseglia, and R.~D. Braatz, ``Constrained zonotopes: A new tool for set-based estimation and fault detection,'' {\em Automatica}, vol.~69, pp.~126--136, 2016.

\bibitem{Bird_HybZono}
T.~J. Bird, H.~C. Pangborn, N.~Jain, and J.~P. Koeln, ``Hybrid zonotopes: A new set representation for reachability analysis of mixed logical dynamical systems,'' {\em Automatica}, vol.~154, p.~111107, 2023.

\bibitem{BIRDthesis_2022}
T.~Bird, ``Hybrid zonotopes: A mixed-integer set representation for the analysis of hybrid systems.'' Purdue University Graduate School, 2022.

\bibitem{Koeln2024_zonoLAB}
J.~Koeln, T.~J. Bird, J.~Siefert, J.~Ruths, H.~C. Pangborn, and N.~Jain, ``{zonoLAB: A MATLAB} toolbox for set-based control systems analysis using hybrid zonotopes.'' American Control Conference, 2024.

\bibitem{Sherali1992}
H.~D. Sherali and C.~H. Tuncbilek, ``A global optimization algorithm for polynomial programming problems using a reformulation-linearization technique,'' {\em Journal of Global Optimization}, vol.~2, pp.~101--112, 1992.

\end{thebibliography}
